%% file: draft_spline.tex
\documentclass[10pt, conference, compsocconf]{IEEEtran}

\usepackage{ifpdf}
\usepackage{cite}

\usepackage[cmex10]{amsmath}
\usepackage{algorithmic}
\usepackage{array}
\usepackage[caption=false,font=footnotesize]{subfig}

\usepackage{fixltx2e}
\usepackage{listings}
\usepackage{algorithmic}
\usepackage{xspace}
\usepackage{parcolumns}

\usepackage{url}

  \usepackage[pdftex]{graphicx}

\usepackage{mdwmath}

\usepackage{amsmath}
\usepackage{algorithmic}
\usepackage{array}

\usepackage{fixltx2e}
\usepackage{color}
\usepackage{listings}
\usepackage{algorithm}
\usepackage{algorithmic}
\usepackage{xspace}
\usepackage{parcolumns}

\usepackage{url}

\newcommand{\code}[1]{{\small\texttt{#1}}}
\newcommand{\trash}[1]{}

\newcommand{\registered}{\textsuperscript{\textregistered}\xspace}
\newcommand{\trademark}{\texttrademark\xspace}

\newcommand{\xphiTM}[0]{Intel\registered Xeon Phi\trademark\xspace}

\newcommand{\xeonsTM}[0]{Intel\registered Xeon\registered processors\xspace}
\newcommand{\xeonTM}[0]{Intel\registered Xeon\registered processor\xspace}
\newcommand{\xeonRR}[0]{Intel\registered Xeon\registered}

\newcommand{\vtuneTM}{Intel\registered VTune\trademark Amplifier 2016\xspace }
\newcommand{\compilerReg}{Intel\registered C++ Compilers\xspace }

\newcommand{\VGH}{\code{VGH}}
\newcommand{\VGL}{\code{VGL}}
\newcommand{\V}{\code{V}}

\newcommand{\g}{\code{g}}
\newcommand{\h}{\code{h}}
\newcommand{\N}{\code{N}}

\newcommand{\XCommand}{\code{X}}

\newcommand{\z}{\code{z}}

\newcommand{\pragmaSIMD}{\code{\#pragma omp simd }}

\newcommand{\BsplineSoA}{\code{BsplineSoA }}
\newcommand{\WalkerSoA}{\code{WalkerSoA }}

\newcommand{\CompilerDisclaimer}{
Optimization Notice: Intel's compilers may or may not optimize to the same degree for non-Intel microprocessors for optimizations that are not unique to Intel microprocessors. These optimizations include SSE2, SSE3, and SSSE3 instruction sets and other optimizations. Intel does not guarantee the availability, functionality, or effectiveness of any optimization on microprocessors not manufactured by Intel. Microprocessor-dependent optimizations in this product are intended for use with Intel microprocessors. Certain optimizations not specific to Intel microarchitecture are reserved for Intel microprocessors. Please refer to the applicable product User and Reference Guides for more information regarding the specific instruction sets covered by this notice.
}

\newcommand{\BenchmarkDisclaimer}{
Software and workloads used in performance tests may have been optimized for performance only on Intel microprocessors. 
Performance tests, such as SYSmark and MobileMark, are measured using specific computer systems, components, software, operations and functions. Any change to any of those factors may cause the results to vary. You should consult other information and performance tests to assist you in fully evaluating your contemplated purchases, including the performance of that product when combined with other products.   For more complete information visit www.intel.com/benchmarks.  
\\
Intel, Xeon, and Intel Xeon Phi are trademarks of Intel Corporation in the U.S. and/or other countries.
}

\begin{document}

%

\title{Optimization and parallelization of B-spline based orbital evaluations in QMC 
on multi/many-core shared memory processors}

%
\author{\IEEEauthorblockN{Amrita Mathuriya\IEEEauthorrefmark{1}\IEEEauthorrefmark{4},
Ye Luo\IEEEauthorrefmark{2},
Anouar Benali\IEEEauthorrefmark{2},
Luke Shulenburger,\IEEEauthorrefmark{3} and
Jeongnim Kim\IEEEauthorrefmark{1} 
}
\IEEEauthorblockA{\IEEEauthorrefmark{1} Intel Corporation}
\IEEEauthorblockA{\IEEEauthorrefmark{2} Argonne National Laboratory}
\IEEEauthorblockA{\IEEEauthorrefmark{3} Sandia National Laboratories}
\IEEEauthorblockA{\IEEEauthorrefmark{4} \code{Corresponding author:} \code{amrita.mathuriya@intel.com}
}}

\maketitle

\begin{abstract}
\input abstract
\end{abstract}


%
\IEEEpeerreviewmaketitle

\input intro

\input bsplineintro

\input profile
\input method

\input results

\input conclusion

\section*{Acknowledgment}
\noindent
We would like to thank Lawrence Meadows, John Pennycook, Jason Sewall, Harald Servat 
and Victor Lee for their helpful discussions and reviewing this manuscript.  This
work is supported by Intel Corporation  to establish the
Intel Parallel Computing Center at Argonne National Laboratory.  LS was
supported by the Advanced Simulation and Computing - Physics and Engineering
models program at Sandia National Laboratories.  AB was supported through the
Predictive Theory and Modeling for Materials and Chemical Science program by
the Office of Basic Energy Science (BES), Department of Energy (DOE).  Sandia
National Laboratories is a multi- program laboratory managed and operated by
Sandia Corporation, a wholly owned subsidiary of Lockheed Martin Corporation,
for the U.S. Department of Energy’s National Nuclear Security Administration
under Contract No. DE-AC04-94AL85000. This research used resources of the
Argonne Leadership Computing Facility, which is a DOE Office of Science User
Facility supported under Contract No. DE-AC02-06CH11357.


\bibliographystyle{IEEEtran}
\bibliography{qmcpack}

\end{document}

%% file: abstract.tex
B-spline based orbital representations are widely used in Quantum Monte Carlo (QMC) simulations of
solids, historically taking as much as 50\% of the total run time.  Random accesses to a
large four-dimensional array make it challenging to efficiently utilize caches
and wide vector units of modern CPUs.  We present node-level optimizations of
B-spline evaluations on multi/many-core shared memory processors.  To increase
SIMD efficiency and bandwidth utilization, we first apply data layout
transformation from array-of-structures to structure-of-arrays (SoA).  Then by
blocking SoA objects, we optimize cache reuse and get sustained
throughput for a range of problem sizes.  We implement efficient nested
threading in B-spline orbital evaluation kernels, paving the way towards
enabling strong scaling of QMC simulations.  These optimizations
are portable on four distinct cache-coherent architectures and result in up to
5.6x performance enhancements on \xphiTM processor 7250P (KNL), 5.7x on \xphiTM coprocessor 7120P, 
10x on an \xeonTM E5v4 CPU and ~9.5x on BlueGene/Q processor.  
Our nested threading
implementation shows nearly ideal parallel efficiency on KNL up to 16 threads.
We employ roofline performance analysis to model the impacts of our
optimizations.  This work combined with our current efforts of optimizing other
QMC kernels, result in greater than 4.5x speedup of miniQMC on KNL.


%% file: intro.tex
\section{Introduction}
\begin{figure}[b]
\centering
\centering
\includegraphics[scale=0.3]{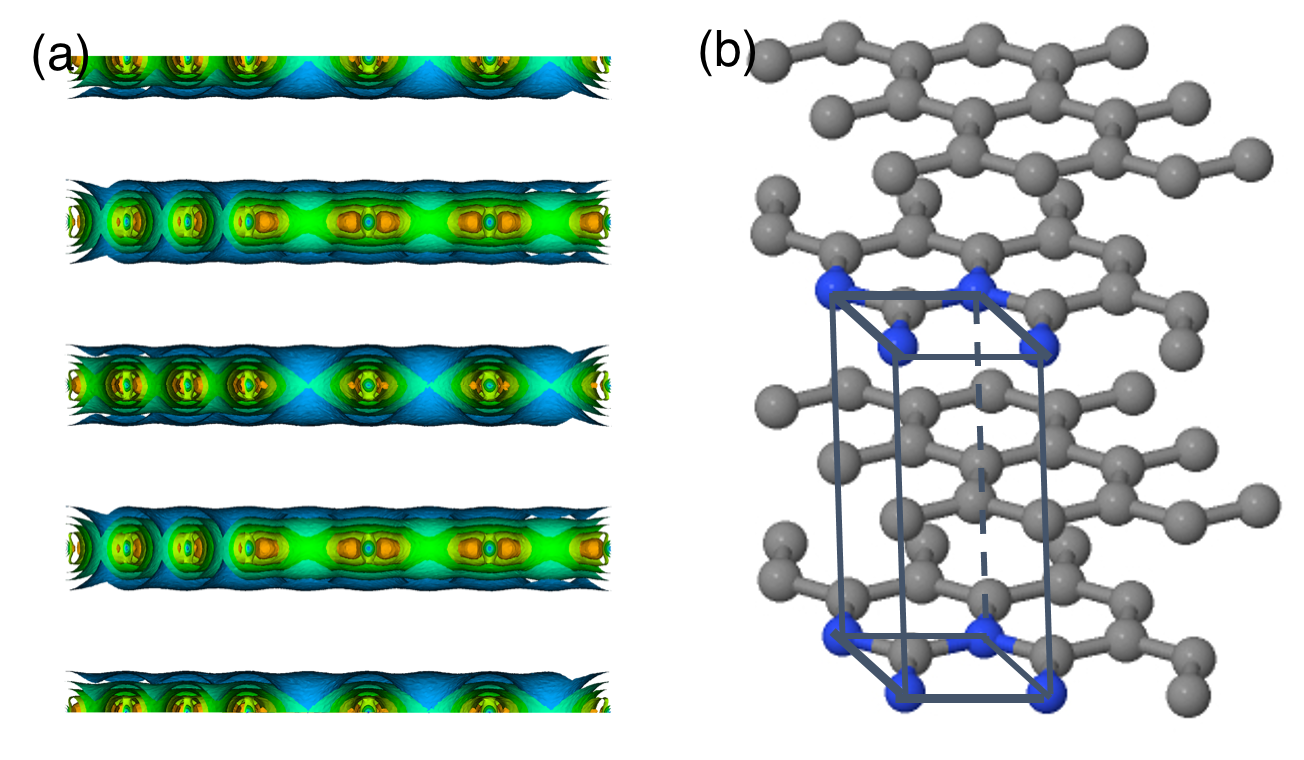}
\caption{(a) DMC charge-density of AB stacked graphite ~\cite{graphiteQMC} and
(b) the ball-and-stick rendering and the 4-Carbon unit cell in blue.}
\label{fig:graphite}
\end{figure}

\noindent
With the advances in algorithms and growing computing powers, quantum Monte Carlo
(QMC) methods have become a leading contender for high accuracy calculations
for the electronic structure of realistic systems. It is general and applicable
to a wide range of physical and chemical systems in any dimension, boundary
conditions etc. The favorable scaling of $\mathcal{O}(N^3)$ 
for an $N$-electron system and ample
opportunities for parallelization make QMC methods uniquely powerful tools to
study the electronic structure of realistic systems on large-scale parallel
computers.  
Despite the high computational cost of the
method, this scalability has enabled calculations on a wide variety of
systems, including large molecules\cite{ellipticene}, layered
materials such as the graphite shown in Figure~\ref{fig:graphite}\cite{graphiteQMC, intercolationQMC, phosphorus}, transition
metal oxides\cite{zinc, cuprate} and general bulk properties of solids\cite{benchmark}.

%
%

A prime trend 
in HPC field is the increase in parallelism available
on a single SMP node. The average number of nodes of big clusters has not
increased in recent years but the compute capacity of a node has been 
increasing through more cores, multiple hardware threads per core, and wider
SIMD units.  The many-core architecture expands the parallelism on a node
dramatically. For example, the second generation \xphiTM processors, formerly
codenamed Knights Landing (KNL)~\cite{knlhc15}, have up to 72 cores, 4 threads
per core and a double-precision SIMD width of 8. Its theoretical peak is
more than 10 times of one IBM Blue Gene/Q (BG/Q) node of 16 cores, 4 threads per
core and a double-precision SIMD width of 4~\cite{bgq}.  
To take full advantage of modern day systems, it is essential to 
utilize SIMD units and cache efficiently. 

One of the main computational bottlenecks in QMC simulations is the evaluation
of $N$ single particle orbitals (SPOs) for an $N$-electron system.  A B-spline
basis is the most efficient representation for SPOs whose cost per orbital is
$\mathcal{O}(1)$.  
Spline interpolations are widely used in many applications including
classical and quantum molecular dynamics\cite{deserno1998mesh} to lower the
cost of complex analytic function evaluations.
In QMC, B-spline based orbitals provide orders of magnitude better
time-to-solution than other standard basis sets such as Gaussian-type orbitals
even for a modest problem size for the same accuracy.
This work presents processes to improve on-node performance of 
B-spline evaluation routines in QMCPACK~\cite{qmcpack}.
The random access
patterns that are inherent in QMC make achieving high performance of
B-spline based SPOs on any architecture challenging. 
Its arithmetic intensity, FLOPS
per bytes, is low and its performance is sensitive to the memory
subsystems -- bandwidths and cache levels and their sizes.  
We demonstrate how one can achieve  
efficient vectorization, optimally utilize the multi-level cache system 
and exploit a large number of
hardware threads as a proof of concept. 

\textbf{SIMD and cache utilization}:
Extensive analysis of QMCPACK simulations reveals that SIMD efficiency is low
except for B-spline evaluations in intrinsics and BLAS/LAPACK routines
that are available in optimized libraries, such as Intel MKL~\cite{mklCite}. 
The abstractions for 3D physics, defined in array-of-structures (AoS) formats, 
such as representing positions for $N$ particles with 
\code{R[N][3]} data layout,
are primarily responsible
for the low SIMD efficiency.
AoS representations for physical entities are logical
for expressing concepts 
and have been widely adopted in HPC applications.  
However, the computations using them 
are not efficient on modern CPUs.
Implementing ``hotspots'' using architecture-specific intrinsics is possible but it is
not scalable nor portable. 
Also currently, without any cache-blocking optimizations,
the working set fall out of cache for large problems, 
causing low cache uitlization.
Our goal is to transform all QMCPACK core routines to increase
on-node performance and science productivity, while minimizing the development
cost, and this work presents the first most important step to achieve the goal. 

\textbf{Beyond walker parallelism}:
The common parallelization strategy in QMC is distributing walkers (Monte Carlo
samples) among $N_p$ tasks, where $N_p$ is the number of parallel processing
units for a simulation.
The OpenMP/MPI hybrid programming model adopted in QMCPACK has proven to be
very effective on clusters of multi-core shared-memory processor (SMP) nodes.
It minimizes the memory footprint per MPI task, 
and improves the overall parallel efficiency by considerably
reducing collective communication time at a large task
count~\cite{qmcpackhybrid}.
However, the cost of processing each walker increases as $\mathcal{O}(N^3)$ or
higher for an $N$-electron system. 
Also, the overall memory usage on a node increase as $\mathcal{O}(N_w N^2)$
for $N_w$ walkers.
This makes it essential to parallelize the execution of
each walker 
for reducing the total time to
solution $(T_{tot})$ and memory footprints;
hence, expanding the applicability of QMC methods to
larger system sizes.  We implement efficient nested threading for B-spline based
SPO evaluations, paving the way towards reducing $T_{tot}$ and enabling 
strong scaling of QMC simulations. 




\textbf{Contributions and results}: \label{sec:contributions}
The two main contributions of this work are (i) node-level optimizations to
improve single node efficiency in a portable fashion, without the use of
processor specific intrinsics and (ii) parallelization of the orbital
evaluations for each walker to reduce the time-to-solution in the
strong-scaling runs.  
We develop simplified QMC codes (miniQMC) to facilitate rapid prototyping and
benchmarking.  They capture the computational and data access patterns and have
similar profile of common QMC workloads on a SMP node.  
We demonstrate the impact of our optimizations on four
distinct cache-coherent architectures with extensive analysis and on a wide
range of problems, starting from 64-carbon (128 SPOs) to 2048-carbon
(4096 SPOs) systems.  Finally, we employ the roofline model analysis to quantify the
relative and absolute performance impact from the optimizations.

The key accomplishments of this work are as follows.  We apply data layout
transformation from array-of-structures to structure-of-arrays (SoA) to
increase SIMD efficiency and bandwidth utilization.  We implement ``tunable
tiling'' or AoSoA transformation to achieve efficient cache
utilization allowing sustained throughput across problem sizes.  The obtained
optimal tile size is independent of the problem size and can be tuned once for a
specific architecture with miniQMC.  These optimizations result in up to
5.6x(KNL), 5.7x(KNC), 10x(BDW) and 9.5x (BG/Q) speedups of B-spline routines.  
We provide an efficient nested
threading implementation for each walker (Monte Carlo unit)
and demonstrate more than 14x reduction in the time-to-solution on 16 KNL
nodes.

The optimizations discussed in this publication 
are broadly applicable to many HPC applications for increasing
the utilization of wide SIMD units and caches.
In addition to B-spline kernels, 
we also implement them in miniQMC for optimizing 
the other time consuming QMC kernels.
With the combined work, we obtain greater than 4.5x speedup of miniQMC 
on KNL and BDW processors.

To faciliate easy transition of performance gains proven in miniQMC
to QMCPACK, we take help of object oriented programming. 
We design and implement SoA container classes for particle abstractions in 3D
and develop classes 
hiding the low-level computational engines for SIMD and memory/cache
optimizations from the application-level classes. 
The optimized kernels of miniQMC will directly be called 
from QMCPACK driver routines to transfer the performance gains.



\section{Related work}\label{sec:relatedWork}
\noindent
To conduct cutting edge scientific research on material science, QMCPACK has
been deployed on the current generation of leadership supercomputers, Mira (IBM
Blue Gene/Q) at Argonne National Laboratory and Titan (AMD Opteron CPUs and
NVIDIA Tesla K20 GPU) at Oak Ridge National Laboratory.  
As the most performance-critical component in QMCPACK, 3D B-splines orbitals
have been extensively optimized over the years~\cite{einspline}\cite{qmcpackGA}. 
The highly optimized
routines evaluating B-spline  SPOs are implemented in QPX intrinsics
\cite{YeSC15} on BG/Q, SSE/SSE2 intrinsics on x86  and in CUDA \cite{qmcpackGPU} to
maximize single-node performance.  The QPX and CUDA versions have adopted SoA
data layout
and FMA instructions to achieve high memory bandwidth utilization and FLOP
rates. 
The single precision was first implemented in QMCPACK GPU port with
significant speedups and memory saving and later introduced to the CPU version.
Multiple algorithms using loop transformations and unrolling have been developed.
The QMCPACK build system allows selecting precision and the optimal algorithms and implementations
at compile time.  

The baseline of this work uses the latest optimized
implementations in C/C++ in the official QMCPACK
distribution~\cite{qmcpack,YeSC15}.
Extensive studies in the data transformations to maximize SIMD efficiency on
\xphiTM (co)processors and nested parallelisms are
available~\cite{Pearls1.book,Pearls2.book,stdl,sbb}.

%% file: bsplineintro.tex
\section{QMC and B-spline-based SPOs}
\noindent
In quantum mechanics, all physically observable
quantities for a system containing $N_{el}$ particles can be computed from
the $3N_{el}$-dimensional {\em wave function},
$\Psi(\mathbf{r}_1,\dots,\mathbf{r}_{N_{el}})$.  
The Slater-Jastrow trial wave functions $\Psi_T$, commonly used
in QMC applications for solving electronic structures, are the product of Slater determinants of
single-particle orbitals (SPOs) and a Jastrow factor:
\begin{equation}
\Psi_T= \exp (J)D^{\uparrow} (\{\phi\}) D^{\downarrow} (\{\phi\}),
\label{eq:sj}
\end{equation}
with $N_{el}=N^{\uparrow}+N^{\downarrow}$ for $\uparrow\downarrow$ spins.
For the rest of the paper, we assume  $N_{el}=2N$, $D^{\downarrow}=D^{\uparrow}$.
Here, $\{\phi\}$ denotes a set of SPOs and
\begin{equation}
D = \det[\mathcal{A}]=\det\left|\,
\begin{matrix}
 \phi_1({\bf r}_{1}) & \cdots & \phi_1({\bf r}_{N}) \\
  \,\vdots\hfill&\,\vdots\hfill&\,\vdots\hfill\\
    \phi_{N}({\bf r}_{1}) & \ldots & \phi_{N}({\bf r}_{N})\\
    \end{matrix}
    \right|,
\end{equation}
which ensures the antisymmetric property of a Fermionic wave function upon
exchange of a pair of electrons. The Jastrow factors, which are factorized for
computational efficiency, describe the dynamic correlation, whereas static
correlation is described by the Slater determinants. 

%

In the diffusion Monte Carlo algorithm (DMC), an ensemble of walkers ({\em
population}) is propagated stochastically from generation to generation, where
each walker is represented by $\mathbf{R}$.  In each propagation step, the
walkers are moved through position space by (i) a {\em drift-diffusion
process}. Once a new configuration is sampled, the physical quantities
(observables) such as the kinetic energy and Coulomb potential energies are
computed for each walker, (ii) a {\em measurement stage}.  Finally, the
ensemble averages of the observables and trial energy $E_T$ are computed and depending on
the energy of a walker relative to $E_T$, each walker may reproduce itself, be
killed, or remain unchanged by a (iii) {\em branching process}.

The particle-by-particle moves we employ change only one column of the
$\mathcal{A}$ matrices at a time and the ratio can be computed as
\begin{equation}
\frac{det[\mathcal{A}^{'}]}{det[\mathcal{A}]}=\sum_{n}^{N}\phi_n ({\bf r}_e) * \mathcal{A}^{-1}(n,e) .  
\end{equation}
When the move is accepted, we employ a rank-1 update of $\mathcal{A}^{-1}$
using the Sherman-Morrison formula.  This allows the inverse to be updated in
$\mathcal{O}(N^2)$ time rather than $\mathcal{O}(N^3)$ time for a full
inversion and the ratios in $\mathcal{O}(N)$.  The computations of 
the many-body gradients and laplacian use the same ratio formula as
\begin{equation}
\frac{\nabla{det[\mathcal{A}^{'}]}}{det[\mathcal{A}]}=\sum_{n}^{N} \nabla \phi_n ({\bf r}_e) * \mathcal{A}^{-1}(n,e).  
\end{equation}

\begin{figure}
\centering
\includegraphics[scale=0.375]{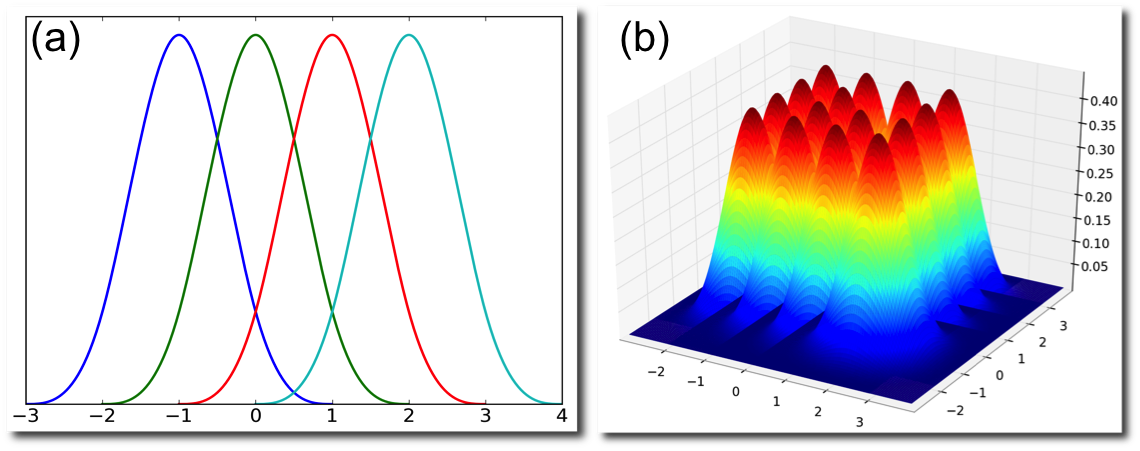}
\caption{The piecewise cubic polynomial basis functions (a) 1D and (b) 2D.}
\label{fig:bbasis}
\end{figure}

The cost to compute the value of a $\phi$ scales linearly
with the number of basis function evaluations which tends to grow with the
system size.  This amounts to $\mathcal{O}(N^2)$ cost for each particle move
and in total SPO evaluations scale as $\mathcal{O}(N^3)$ per Monte Carlo step. 
For this reason, it is efficient to use a localized basis with compact
support.  In particular, 3D tricubic B-splines provide a basis in which only
64 elements are nonzero at any given point in space~\cite{blips4QMC,
einspline} and have complexity of $\mathcal{O}(1)$.
The one-dimensional cubic B-spline is given by,
\begin{equation}
f(x) = \sum_{i'=i-1}^{i+2} b^{i'\!,3}(x)\,\,  p_{i'},
\label{eq:SplineFunc}
\end{equation}
where $b^{i,3}(x)$ are the piecewise cubic polynomial basis functions and
$i=\text{floor}(\Delta^{-1} x)$, the lower bound of $x$ for a grid spacing $\Delta$.
Figure~\ref{fig:bbasis}(a) shows the four basis functions contributing
to the values at $0 \leq x < 1$ when $\Delta=1$.
Constructing a tensor product in each Cartesian
direction, we can represent a 3D orbital as
\begin{equation}
  \phi_n(x,y,z) = 
  \!\!\!\!\sum_{i'=i-1}^{i+2} \!\! b_x^{i'\!,3}(x) 
  \!\!\!\!\sum_{j'=j-1}^{j+2} \!\! b_y^{j'\!,3}(y) 
  \!\!\!\!\sum_{k'=k-1}^{k+2} \!\! b_z^{k'\!,3}(z) \,\, p_{i', j', k',n}.
\label{eq:TricubicValue}
\end{equation}
This allows for rapid evaluation of each orbital in constant time.
Furthermore, the basis is systematically improvable with a single spacing
parameter, so that the accuracy is not compromised.  The coefficients $\{p\}$
are the interpolation tables for each orbital and remain constant throughout
the simulations.

The use of 3D tricubic B-spline SPOs greatly improves the time-to-solution.
For a typical modest example problem with 32 electrons, the speed up of
B-spline is more than six-fold over an equivalent Plane-Wave (PW) basis set. The
advantage of the B-spline SPOs grows as the system size grows.  
This computational efficiency comes at the expense of increased
memory use.
To keep the memory footprints small, QMCPACK uses 
hybrid parallelism with OpenMP/MPI where all the threads share
the read only coefficient table.



%% file: profile.tex
\section{Baseline and miniQMC}
\noindent
We use the CORAL benchmark 4x4x1 problem~\cite{coral} to establish the baseline
performance.
This benchmark represents typical QMC workloads on
current generation of HPC systems. It solves 256 electrons of 64-atom AB-stacked graphite
system consisting of 4 by 4 periodic images of the 4-atom unit cell, shown
in blue in Fig.~\ref{fig:graphite}(b). It defines the grid sizes
$N_x$=$N_y$=48 and $N_z$=60 of $N$=128 orbitals.
The details
of the systems used are listed in Table~\ref{tab:sys}. Systems
are described in section~\ref{results}, containing the experiments.
\compilerReg\footnote{\CompilerDisclaimer} version 16 update 2 are used on 
Intel platforms~\cite{intelcomp} and IBM XL C/C++ 12.1 is used on BG/Q~\cite{ibmcomp}.
Also, we use \vtuneTM (VTune) \cite{vtuneCite} advanced hotspot profiling on Intel platforms and HPCToolkit~\cite{hpctoolkit} on BG/Q for run time estimation. 
Hyperthreading is beneficial and is used for the runs in
Table~\ref{tab:FullProfile} and the data presented in this work.


\begin{table}[!t]
\caption{System configurations.}
\label{tab:sys}
\begin{minipage}{0.45\textwidth}
\centering
\begin{tabular}{|l|c|c|c|c|}
\hline
& BDW & KNC & KNL & BG/Q\\
\hline
Processor & E5-2697v4 & 7120P & 7250P & PowerPC A2\\
\hline
\# of cores & 18 & 61 & 68 & 17 \\
\hline
Hyperthreading  & 2 & 4 & 4 & 4 \\
\hline
SIMD width(bits) & 256 & 512 & 512 & 256\\
\hline
Freq. (GHz) & 2.3 & 1.238 & 1.4 & 1.6 \\
\hline
\hline
L1(data) & 32 KB & 32 KB  & 32 KB & 16 KB \\
\hline
L2 (per core) & 256 KB & 512 KB  & 1 MB  & \\
              &        &         & \scriptsize{Per tile} & \\
\hline
LLC (shared) & 45 MB & & &32 MB \\
\hline
Stream BW(GB/s)& 64 & 177 & 490 & 28 \\
\hline
\end{tabular}
\end{minipage}
\end{table}

\begin{table}[b]
    \caption{Single node run time profile in \% of the 
      4x4x1 
      CORAL benchmark for publicly available QMCPACK
    }
    \label{tab:FullProfile}
    \centering
    \begin{tabular}{|c|r|r|r|r|}
    \hline
   & BDW & KNC & KNL& BG/Q  \\
    \hline
    B-splines& 18 & 28 & 21 & 22\\
    \hline
    Distance Tables & 30 & 23 & 34 & 39\\
    \hline
    Jastrow & 13 & 19 & 19 & 21\\
    \hline
    \hline
    \# cores used  & 18 & 60 & 64 & 16\\
    \hline    
    OpenMP threads  & 36 & 240 & 256 & 64\\
    \hline
    \end{tabular}
\end{table}

\begin{table}[b]
    \caption{
      Profile for miniQMC (equivalent to table \ref{tab:FullProfile}) 
      with the optimized Distance-Tables and Jastrow kernels.
    }
    \label{tab:ProfileminiQMC}
    \centering
    \begin{tabular}{|c|c|c|c|}
    \hline
    & B-splines & Distance Tables  & Jastrow \\
    \hline
    KNL & 68.5 & 20.3 & 11.2 \\
    \hline
    \xeonRR E5-2698 v4 & 55.3 & 22.6 & 22.1 \\
    \hline
    \end{tabular}
\end{table}

Table~\ref{tab:FullProfile} shows the profile of the benchmark 
with publicly released QMCPACK~\cite{qmcpack} on single nodes. 
The
main computational groups are B-splines, distance tables and one-body and
two-body Jastrow evaluations.  Their total amounts to $60\%$-$80\%$ across
the platforms.  
Rest of the time is mostly spent on 
 the assembly of SPOs
using B-spline outputs, determinant updates and inverses. 
The optimization steps mentioned in
Sec.~\ref{sec:relatedWork} and QPX specializations accounted to reduce 
the B-spline 
share of the profile from 22\% to 11\% on BG/Q.

In parallel optimization efforts, we optimize Distance-Tables and Jastrow kernels 
with the SoA transformation of particle position abstractions 
and with other algorithmic improvements. 
With these,
B-spline routines consume more than 55\%
of run time for miniQMC as shown in table~\ref{tab:ProfileminiQMC}.

%% file: method.tex
\begin{figure}
\centering
\begin{minipage}{.42\textwidth}
\lstset{language=c++,
  basicstyle=\ttfamily\scriptsize,
  commentstyle=\itshape\color{blue},
  otherkeywords={bspline, V, VGL, VGH},
  breaklines=true,
  numbers=left,
  frame=single,
  captionpos=b
}
\begin{lstlisting}
//Number of splines = N, iterations = niters
//random samples = ns = 512 
//dimensions of 3D position grid (nx, ny, nz)

using Pos3=float[3];
class WalkerAoS {T v[N], g[3*N], l[N], h[9*N];};

//Create and init read only 4D table P once.
BsplineAoS bSpline(nx, ny, nz, N);

//Create and run walkers in parallel. 
#pragma omp parallel 
{
  //Contains private copy of outputs.
  WalkerAoS w;

  // generate random samples for testing.
  Pos3 vPos[ns], vglPos[ns], vghPos[ns];
  generateRandomPos(vPos, vglPos, vghPos, ns );

  for(int i=0; i<niters; i++){
    for ( int j=0; j < ns; j++ )
      bSpline.V(vPos[j],w.v);
    for ( int j=0; j < ns; j++ )
      bSpline.VGL(vglPos[j],w.v,w.g,w.l);
    for ( int j=0; j < ns; j++ )
      bSpline.VGH(vghPos[j],w.v,w.g,w.h);
  } 
} 
\end{lstlisting}
\end{minipage}
\caption{Simplified miniQMC only containing B-spline evaluations. \label{alg:bspline_thd}}
\end{figure}

\begin{figure*}[thb]
\begin{minipage}{.48\textwidth}
\lstset{language=c++,
  basicstyle=\ttfamily\scriptsize,
  commentstyle=\itshape\color{blue},
  keywordstyle=\bf{\color{red}},
  deletekeywords={class,void,int,const},
  otherkeywords={VGH,3*n+0,3*n+1,9*n+0,9*n+1,p[n],(a)},
  breaklines=true,
  captionpos=b,
  label={fig:baselineCode}
}
\begin{lstlisting}[frame=tlrb]
(a) class BsplineAoS {
T P[Nx][Ny][Nz][N]; // read only and shared among threads. 
void VGH(T x, T y, T z, T* v, T* g, T* h) {
  //compute the lower-bound index i0,j0,k0
  //compute prefactors using (x-x0,y-y0,z-z0)

  for(int i=0; i<4; ++i)
    for(int j=0; j<4; ++j)
      for(int k=0; k<4; ++k) {
        const T* p=P[i+i0][j+j0][k+k0];
        #pragma omp simd
        for(int n=0; n<N; ++n) {
         v[n]  += F(p[n]);
         g[3*n+0]+=Gx( p[n]);g[3*n+1]+= Gy( p[n]);
         h[9*n+0]+=Hxx(p[n]);h[9*n+1]+= Hxy(p[n]);
         ...  }}}
};
\end{lstlisting}
\end{minipage}\hfill
\begin{minipage}{.48\textwidth}
\lstset{language=c++,
  basicstyle=\ttfamily\scriptsize,
  commentstyle=\itshape\color{blue},
  keywordstyle=\bf{\color{red}},
  deletekeywords={class,void,int,const},
  otherkeywords={VGH,p[n],(b)},
  breaklines=true,
  captionpos=b,
  label={code:soa}
}
\begin{lstlisting}[frame=tlrb]
(b) class BsplineSoA {
T P[Nx][Ny][Nz][N]; // read only and shared among threads. 
void VGH(T x, T y, T z, T* v, T* g, T* h) {
  //compute the lower-bound index i0,j0,k0
  //compute prefactors using (x-x0,y-y0,z-z0)
  T *gx=g, *gy=g+N, *gz=g+2*N;
  T *hxx=h,*hxy=h+N,*hxz=h+2*N, 
    *hyy=h+3*N,*hyz=h+4*N,*hzz=h+5*N;
  for(int i=0; i<4; ++i)
    for(int j=0; j<4; ++j)
      for(int k=0; k<4; ++k) {
        const T* p=P[i+i0][j+j0][k+k0];
        #pragma omp simd
        for(int n=0; n<N; ++n) {
         v[n]+= F(p[n]);
         gx[n] += Gx(p[n]);   gy[n]+= Gy(p[n]);
         hxx[n]+= Hxx(p[n]); hxy[n]+= Hxy(p[n]);
         ...  }}}
};
\end{lstlisting}
\end{minipage}
\caption{\VGH\; using the Gradients and Hessians (a) in AoS memory layout
and (b) in SoA memory layout.}
\label{fig:baselineCode}
\end{figure*}

Figure~\ref{alg:bspline_thd} shows the pseudocode for 
simplified miniQMC 
that captures the essential computations involving B-spline evaluations.
An object named bSpline, of BsplineAoS class 
is created at L10, which creates and initializes 
4D coefficient array \code{P} with dimensions \code{[nx][ny][nz][N]}.
This read only \code{P} array is shared by all the threads through bSpline object.
Independent Monte Carlo units called walkers, are created
using \code{pragma omp parallel} at L13.
To imitate the random access nature of QMC, 
each walker generates 
\code{ns} random positions \code{(x,y,z)} (shown on L18) 
for each of the three B-spline kernels, namely \V, \VGL\;
and \VGH.
Each walker 
creates an object \code{w} of \code{WalkerAoS} class (shown on L16) 
which contains its own copy of output arrays. 
With the generated random positions, each walker fills up 
these output arrays.
The iteration loop at L20 represents working through multiple generations of
Monte Carlo simulation.


The pseudocode in Fig.~\ref{fig:baselineCode} shows the kernel of
\VGH\; which computes $N$ values, gradients (real 3D vectors) and Hessians
(symmetric 3D tensors). A total of 13$N$ output components are evaluated. 
The input  position \code{(x,y,z)} is random and provided by the miniQMC
driver (Fig.~\ref{alg:bspline_thd} at L22, L24 and L26) to mimic QMC 
random moves by the quantum forces.
The outputs \code{v,g} and \h\;
passed as parameters to the function call, are the starting addresses of the
particle attributes,
\code{V[N]} (values) and AoS types of \code{G[N][3]} (gradients)  and
\code{H[N][3][3]} (Hessians). 
Gradients (Hessians) are the first (second)
derivatives with respect to the particle position and are used to compute
quantum forces and Laplacians. 
The allocation of the $P$ coefficient array is done as 1D array and uses
an aligned allocator 
and includes padding to ensure the alignment of \code{P[i][j][k]} to a
512-bit cache-line boundary. 
All the computations in miniQMC are performed in single precision.  

In 4D memory space, accesses to $P$ start at a random input grid point
\code{P[i0][j0][k0]} that satisfies $x_{i0}\le x < x_{i0}+\Delta_x$, for $\Delta_x$
the grid spacing in $x$-direction. The same conditions apply to $y$ and $z$.
Total of 64 input streams are issued to access $N$ coefficient values. In total, 64$N$
stride-one reads and 13$N$ mixed-strided accumulations are executed for each random
input point.  For large problems, 
the arithmetic intensity is low at 1 FMA for each accumulation of the output value
and therefore, memory 
bandwidth plays a critical role in deciding the throughput of the B-spline routines.
The cost of computing $\{b\}$ at \code{(x,y,z)} in 
Eq.~\ref{eq:TricubicValue} is
amortized for $N$, which had big impact on the performance of scalar
processors.

Functions  \V\;
and \VGL\; differ from \VGH\; by how many components are computed, while having
the same computational and data access patterns.  
Based on the simulation cell type, either \VGH\; or \VGL\; is used.
\V\; is used with pseudopotentials for the local energy computation. For the graphite systems,
\VGH\; is used during the drift-diffusion phase. 
Multiple versions of B-spline implementations are available
in QMCPACK and we use the optimized CPU algorithm~\cite{YeSC15} as the
baseline.


\begin{figure}[b]
\centering
\centering
\includegraphics[scale=0.3]{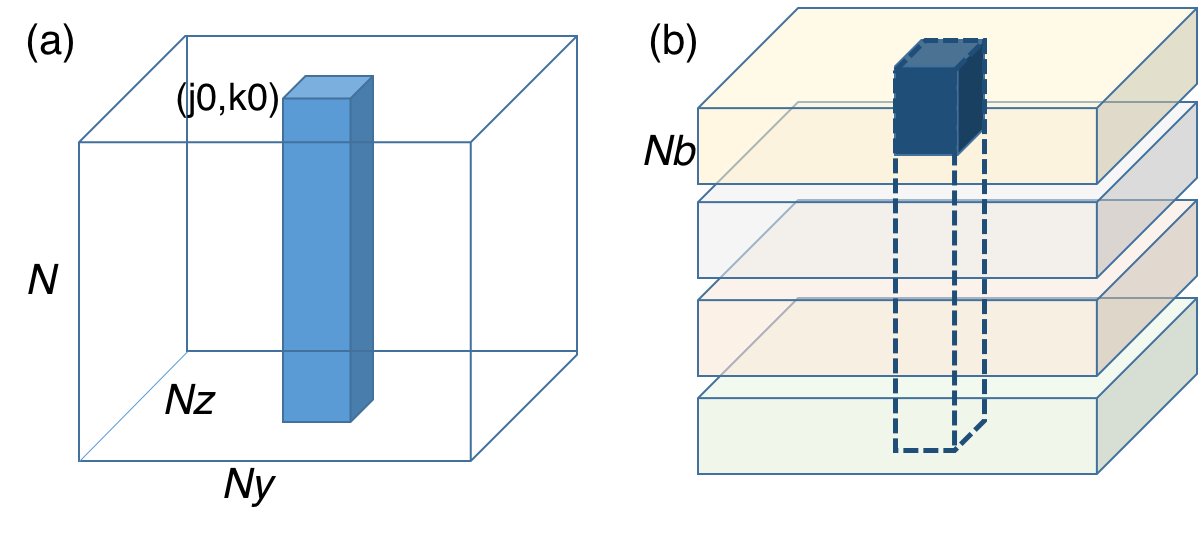}
\caption{Data access pattern of read-only B-spline coefficients $P$ at a random position $(x,y,z)$
and $j0$=floor$(y/\Delta_y)$ \textit{etc.}: (a) current
  einspline library and (b) AoSoA implementation. The outermost $x$ dimension is not shown. 
}
\label{fig:data_access}
\end{figure}

\section {Optimizations and Parallelization }\label{sec:method}
\noindent
In this section, we describe the process and techniques to 
optimize B-spline kernels with SOA and AoSoA data layout transformations.
Further, we describe the nested threading implementation,
which helps reducing time to solution and memory footprints.


\subsection{AoS-to-SoA transformation of outputs}
\noindent
Aligned data objects and contiguous accesses 
for reading and writing 
are essential for efficient vectorization.  
The innermost loop in Fig.~\ref{fig:baselineCode}(a)
runs over number of splines 
$\N$\; and provides sufficient work for efficient 
SIMD parallelism. 
Vectorization of this inner most loop 
results in streamed unit stride accesses
of read only \code{P} array.  
However, accesses to the \g(\h) array have a 3(9)-strided pattern due to
the AoS of the particle abstractions in 3-dimensional space.  
For example, \g\; assumes the internal ordering  of the gradients in
\code{[xyz|xyz|..|xyz]} sequence. 
These strided
accesses result in gather and scatter instructions, which are less efficient
than contiguous loads and stores and reduce SIMD efficiency. 
A better way to store this data is in a
structure-of-arrays (SoA) format and use three separate streams for each
component.  It lets us align the individual output streams for efficient
loading and storing and also eliminates the
need to use gather/scatter instructions.  


Figure~\ref{code:soa} shows the pseudocode with AoS-to-SoA data layout
transformation of the gradient and Hessian arrays.  
\code{BsplineSoA} class replaces \code{BsplineAoS} class 
in this optimized version for Fig.~\ref{alg:bspline_thd}.
This class uses
three separate streams for the gradient vector and 6 streams for the Hessian
tensor.  Only a few components are shown for brevity.  A minor change is made to
exploit the symmetric nature of the Hessian, which results in a total of 10 (1+3+6)
output streams instead of 13 for the baseline. To help the compiler
auto-vectorize and ignore any assumed dependencies, we place \pragmaSIMD
(OpenMP 4.0) on top of the inner most loop.  
Also, we inform the compiler about the alignment of output and
input arrays using directives. 
These changes lead to efficient
vectorization over $N$ and ensure stride-one accesses to both $P$ and
the output arrays.  
In addition to the AoS-to-SoA transformation, we do a few other optimizations 
to the \VGL\; function which are already present for \VGH\;; such as
unrolling the loop over the \z\; dimension, and move the
allocation of temporary arrays of the baseline code out of the loops.

Data-layout transformations from AoS to SoA are commonly employed by HPC
applications on \xphiTM~\cite{Pearls1.book,Pearls2.book} and have been shown to be essential
on the processors with wide SIMD units. 
The same transformation boosts 
performance of the other critical computational steps 
involving distance tables and Jastrow of QMCPACK.
To minimize the impact on theoretical development,
it is important to be able to implement the data layout transformation 
with minimal source code changes. 
To achieve this while keeping the performance penalty low,
we only modify the code 
in performance critical regions 
to explicitly use the SoA containers 
representing abstractions for particle positions,
and
overload their square bracket operators 
to return the particle positions at an index, in the current AoS format. 
This lets us keep the internal data layout in SoA format and allows
the use in both AoS and SoA formats. 
Other HPC codes written with object oriented programming
can take advantage of this technique to gain performance 
with minimal programming efforts. 

%

\subsection{AoSoA transformation (``tiling")}
\noindent
Efficient cache utilization is critical in getting high performance on large
cache-based architectures.  
To realize this, tiling or blocking optimizations 
have proven to be highly effective
in many grid-based applications~\cite{Pearls1.book}.
This is also probably one of the most difficult and a tricky one to implement 
in a way that allows maximizing performance. 
We present our method of tiling proven to be efficient for B-spline
routines in QMC.

Simulating periodic images of a primitive unit cell for the SPOs
involves keeping the grid $N_g=(nx,ny,nz)$ constant and 
increasing the number of splines $N$, for solving larger systems. 
Accesses to the read only $4D$ coefficient array $P$
are random, limiting the cache reuse.  
The working set size in bytes for this single precision (SP)
array is $4N_gN$, making it 
too big to fit in to even
last level caches, for typical $N_g$ values.  
Also, output working set size grows with $N$;
for example, full SP output working set size in bytes for \VGH\; is $40 N N_w$, 
for $N_w$ walkers.
For big N, these arrays fall out of caches.
We apply AoSoA transformation or tiling optimization to increase 
performance by better cache utilization. 

After the SoA transformation,
the spline dimension $N$ becomes the innermost
and contiguous dimension for both the input and output arrays.
We tile both, input 4D coefficient array $P$ and the output arrays,
along the spline dimension $N$.
We create arrays of \code{BsplineSoA} and \code{WalkerSoA} 
objects, effectively splitting the coefficients $P$ and outputs
along $N$ in $M$ groups. We use $N_b=N/M$ as the tile size.
Tiling of $P$ array is pictorially shown 
in Fig.~\ref{fig:data_access}(b). 
Figure \ref{code:tiling} is a pseudo user code using $M$ \BsplineSoA and
\WalkerSoA objects with $N_b$ as the spline dimension.  
Object oriented programming in QMCPACK allows us to
use the 
SoA data containers 
developed for the previous section without any modification.

\begin{figure}[bhtp]
\centering
\begin{minipage}{.42\textwidth}
\lstset{language=c++,
  basicstyle=\ttfamily\scriptsize,
  keywordstyle=\bf{\color{black}},
  commentstyle=\itshape\color{blue},
  numbers=left,
  breaklines=true,
  frame=single
}
\begin{lstlisting}
int Nb = N/M; // tile size.
class WalkerSoA {T v[Nb], g[3*Nb], l[Nb], h[6*Nb]};
// SoA object array containing P[nx][ny][nz][Nb]
BsplineSoA bs[M](Nb); 

// Parallel region for creating walkers.
#pragma omp parallel {
  WalkerSoA w[M](Nb);
#pragma omp parallel for
{
  for(int t=0; t<M; ++t)
    for(int j=0; j<ns; j++)
      bs[t].VGH(vghPos[j], w[t].v, w[t].g, w[t].h);
}}
\end{lstlisting}
\end{minipage}
\caption{A pseudocode with tiling and nested threading.\label{code:tiling}}
\end{figure}

%
%
%
The AoSoA transformation 
enhances spatial and temporal locality by reducing 
working set size.
The \VGH\; working set size
becomes $4N_gN_b$ and $40 N_w N_b$ in bytes for inputs and outputs respectively.
It allows the output 
working set to be kept in cache for accelerated reduction operations
for appropriate $N_b$ values.
Also, the tiled $P$ array can stay in shared LLC (if available)
for small $N_b$ values.
In addition, this AoSoA 
transformation 
exposes parallelism that will be utilized to shorten
computational time in the strong-scaling sense.  


%
\subsection {Exposing Parallelism Within Walker Update}
\noindent
Moving forward, exposing parallelism within a walker 
and increasing scaling to more nodes,
are 
becoming necessary i) to solve problems faster 
and ii) to reduce memory footprints.  
The
AoSoA data layout transformation described in the previous
subsection exposes natural parallelism of B-spline operations. 
Each tile containing \code{BsplineSoA} and \code{WalkerSoA} objects
shown in Fig.~\ref{code:tiling} can be
executed independently, without any synchronization.
The line 9 in Fig.~\ref{code:tiling} shows a 
method for exploiting this parallelism 
using \code{\#pragma omp parallel for}.
The implementation in miniQMC adopts an explicit data partition scheme by assigning $n_{th}$ threads
for each walker and distributing $M$ objects among $n_{th}$ threads.  
This avoids any potential overhead from OpenMP nested run time environment
and sets the maximum performance that can be obtained with 
our cache-and-threading optimizations. 
The pseudocode in Fig.~\ref{code:tiling} is representative of 
the one shown in Fig.~\ref{alg:bspline_thd}, with the tiling and threading optimizations
and omission of redundant details including
the calls to \VGL\; and \V\; routines.

This parallelization strategy
over $M$ independent objects keeps 
the benefits of smaller working sets, enabling efficient
cache utilization.
Also, independent execution of the tiles eliminates
any potential overhead of synchronization.
The working set size in bytes for \VGH\; becomes 
$4N_g N_b n_{th}$ (inputs) and $40N_w N_b n_{th}$ (outputs). 
For the strong scaling runs, we decrease $N_w$ by $n_{th}$ factor,
keeping the output working set size same.
Alternate approach explored for the nested threading includes 
threading over the innermost $N$ dimension, without tiling.
This approach does not reap the benefits of
smaller working sets for efficient cache utilization
and performs worse than the approach chosen here.
We plan to extend this AoSoA design to
parallelize other parts of QMCPACK and are evaluating various parallel
algorithms and run time.  



%% file: results.tex
\section{Experiments and Results}\label{results}
\noindent
Our experiments\footnote{\BenchmarkDisclaimer} are carried out on four different shared memory multi/many-core
processors, 18-core single-socket \xeonTM E5-2697v4 (BDW), the first
generation \xphiTM coprocessor 7120P (KNC), the second generation \xphiTM
processor 7250P (KNL), and IBM Blue Gene/Q (BG/Q). Details of these systems are
shown in Table~\ref{tab:sys}.  As will be discussed, the performance of miniQMC and
QMCPACK are highly sensitive to the memory bandwidth and cache types and cache sizes
and they are deciding  factors for our optimization strategy.  
The computations on KNL are performed in quad cluster mode and flat memory
mode. Memory footprints for all problem sizes are less than 16GB, so we
exclusively use MCDRAM with \code{numactl -m 1} command with the executables.
No code modification is made to manage allocation other than using
the cache-aligned allocators. 

For the analysis, we vary $N$, the number of splines, from 128 
to 4096, from current day problems to large problems planned as the grand-challenge on
pre-exascale systems.
The grid size is kept constant at {\small{$N_g=48\times48\times48$}}, simulating
periodic images of the primitive unit cell for the SPOs, e.g., the blue box in
Fig.~\ref{fig:graphite}(b).  
The tile size $N_b$ is a tunable parameter on each system. We plan
to provide an auto-tuning capability using miniQMC to guide the 
production runs similar to FFTW's solution using wisdom files~\cite{FFTW05}.

We use $T$ for the throughput per node, a QMC specific 
metric \code{(operations/sec)}
to compare performance across problem sizes and processors.  
$T$ represents the
work done on a node per second and is computed as $T_\XCommand=N_w N / t_\XCommand$,
where $t_\XCommand$ is the
total time for $\XCommand=\V,$ $\VGL$\; or $\VGH$. 
It is directly related to the
efficiency of QMC simulations using B-spline SPOs.  
For the ideal performance, $T$ should be independent
of $N$ and the grid sizes $N_g= (nx, ny, nz)$.  Higher $T$ indicates both a more powerful node,
e.g., KNC \textit{vs} KNL, and the higher efficiency of the implementations.
The speedup is the ratio of $T$s before and after optimizations using the same
number of nodes.  As pointed out before, QMCPACK implementation does 
very little communication across nodes 
and the parallel efficiency 
with MPI is excellent. 
This way, we can reliably use the single-node performance for the comparisons.
We choose $N_w$= 36 (BDW), 240 (KNC), 256 (KNL) and 64
(BG/Q), corresponding to 1 walker per thread on each system, as used in current
QMC simulations. 

The rest of the section presents the performance evolution with the processes
detailed in Sec.~\ref{sec:method}.  We focus on the throughput
of \VGH\; 
for the analysis. 
Memory bandwidth usage is a key indicator of the efficiency.
We measure the bandwidth utilization on KNL using VTune.
The performance
characteristics  of \VGL\; is similar to those of \VGH\; on each system and
therefore, only the final data for \VGL\; are presented. 
Kernel \V, having a single floating point output array do 
not need SoA data layout and only benefits with the AoSoA transformation
and the nested threading versions. 


\begin{figure*}
    \centering
    \includegraphics[scale=0.29]{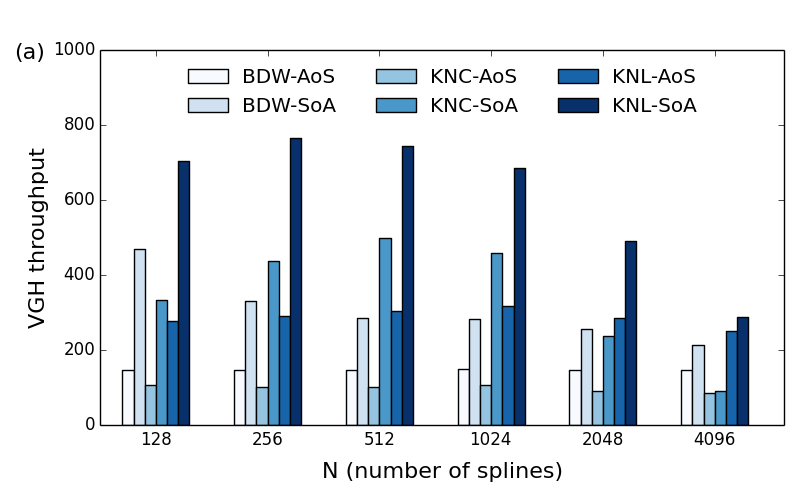}
    \includegraphics[scale=0.29]{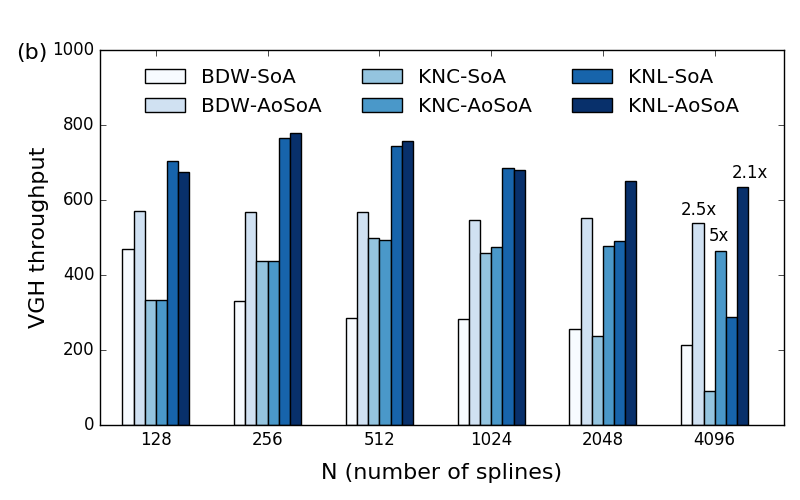}
    \includegraphics[scale=0.29]{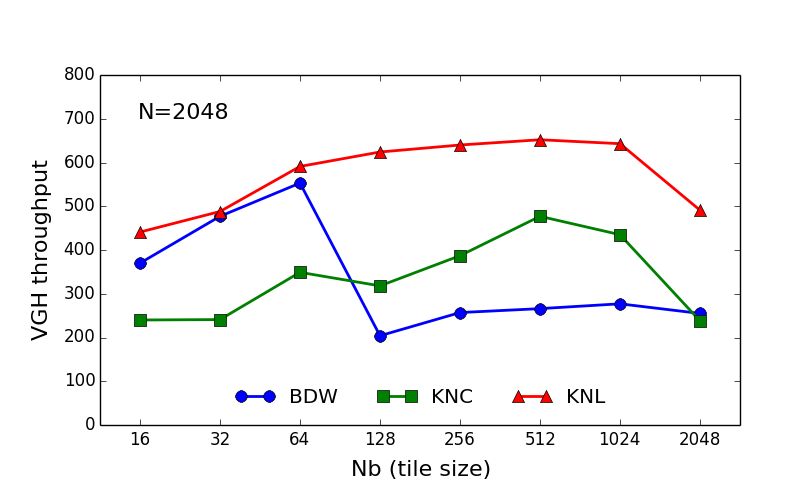}
    \caption{\VGH\; throughput (operations per second), higher the better 
      with (a) AoS-to-SoA and 
    (b) SoA-to-AoSoA (tiling) data layout transformation. (c) 
    Performance of AoSoA at $N=2048$ with $N_b$ tile size, along $N$ (linear dimension).
    }
    \label{fig:all}
\end{figure*}

\subsection{Aos to SoA Data Layout Transformation}
\noindent
Figure \ref{fig:all}(a) presents the performance before and after the AoS-to-SoA
transformation of output arrays, showing 2-4x speedups for small to medium problem
sizes on the Intel processors.  We estimate the current gain 
with vectorization on KNL, 
of an implementation by comparing $T$s 
without auto-vectorization and with vectorization by the compiler. 
We call it as vector efficiency in the following text.
We use
\code{-no-vec -no-simd -no-openmp-simd}
to turn off auto-vectorization by the Intel compilers.  
For a small problem such as 256, the vector efficiency is low at 1.2x  with the
AoS datatypes.  In contrast, this efficiency with SoA objects is greater than
4.  The read bandwidth utilization increases to sustained 238 GB/s
from 60-98 GB/s.  KNC gets the biggest boost with the AoS-to-SoA transformation
being an in-order processor with high memory bandwidth.  This SoA technique is
currently implemented in the QPX  version for BG/Q, leading to 2.2x speedup.  
However, this intrinsics solution is not portable and more importantly, not needed with the transformation.

The AoS-to-SoA transformation achieves good speedups for small to medium
problem sizes but, its effectiveness diminishes as $N$ increases beyond $N$=512.
Almost no speedup is obtained on KNC and KNL at $N$=2048 and 4096.  
The sizes of 
L1/L2 caches are mainly responsible for the poor performance at
large $N$. 
Output arrays fall out of caches for large $N$ values.
This is confirmed with VTune analysis on KNL: the write bandwidth
usage increase to 177 GB/s at $N$=4096 from 38 GB/s at 256.  The drop in $T$
with increase in $N$ is less severe on BDW with the large shared L3 cache. Nevertheless,
the absolute performance goes down for the big problems on all the cache-based systems, we
have tested.  To solve this problem, we apply an AoSoA data layout transformation.

\subsection{AoSoA Data Layout Transformation or Tiling}
\noindent
The Array-of-SoA data layout transformation is a cache-blocking optimization.
We break up the problem into $M$ ``active" working sets to make them 
fit in caches.
Please note that, we divide the working set across the spline dimension $N$,
which is common to both the inputs and outputs.
Output arrays get split and fit in to caches for efficient reduction operations.
Input four dimensional coefficient array $P$ which 
is shared among all the threads,
also gets split across its innermost dimension. 
This allows it 
to fit in the big last level caches.
As shown in Fig.~\ref{code:soa}, access to $P$ array although random 
but, exhibit small amount of spatial locality 
across $x, y$ and $z$ dimensions.
This tiling method also shortens the stride for outer dimensions, 
increasing cache reuse and possibly avoiding page faults
for big grid sizes $N_g$ and large $N$.


Figure~\ref{fig:all}(c) shows $\VGH$ performance with the increasing tile size
for $N$=2048.  
Starting at $N_b=16$, we explore tile sizes in the multiple of two till
$N_b=N$ and call $N_b$ with the highest throughput as the optimal tile size.

A striking feature for BDW is the peak at $N_b=$ 64.
The entire single precision 
working set of roughly 28MB, including
$P$ array 
of $4 N_g N_b$ bytes and the output of $40 N_w N_b$ bytes
fit in the 45MB L3 cache.
The volume
for the input working set at $N_b=128$ of roughly 56MB, exceeds the L3 size 
and therefore, \VGH\; becomes 
memory bandwidth limited for $N_b>64$.
BG/Q having shared L2 of size 32MB, can hold 28MB of working set;
hence also has a peak performance at $N_b=64$.
\xeonsTM with shared LLC behave similarly.

In contrast, KNC and KNL do not show the same pronounced
peak.  
Although L2 caches are kept fully coherent among cores, but 
they are private to cores on KNC and to tiles with two cores on KNL.
Due to random accesses of the large $P$ array, private L2 caches may contain
duplicate copies of elements at any point of time. 
The duplication reduces the effective size for L2 cache. 
Even for $N_b=16$ tile size, the input working set size is 7MB,
which may not allow it to fit in to their L2.
For KNC and KNL, a performance peak is obtained at $N_b=512$.
The improvement comes from fitting output
arrays in cache, allowing efficient reduction operations over 64 grid points.
The performance goes up as $N_b$ increases till $N_b=512$, 
reflecting the amortized cost
of redundant computations of the prefactors for the same random position.


%

Figure~\ref{fig:all}(b) presents $\VGH$ performance before and after the AoSoA 
transformation, showing significant improvement for $N$=2048 and 4096.
We use the obtained optimal $N_b$ on each platform: $N_b$=64
on BDW and  512 on KNC and KNL. 
With tiling, we obtain sustained throughput across the
problem sizes on all the cache-based architectures we have tested. 
On BDW and BG/Q with the shared LLC, $N_b$=64 gives the speedups of 1.2-2.5
(BDW) and 1.2-1.5 (BG/Q) across all the problem sizes.  The speedups on KNC and
KNL are obtained with $N_b$=512. 
With the optimal $N_b$=512 on KNC and KNL,
output arrays stay in L1/L2 caches and the write-bandwidth decreases from 177 GB/s 
to 43 GB/s even for $N$=4096. The vector efficiency 
also increases from 2.5x to 4.2x. 

\begin{figure}[b]
    \centering
    \includegraphics[scale=0.35]{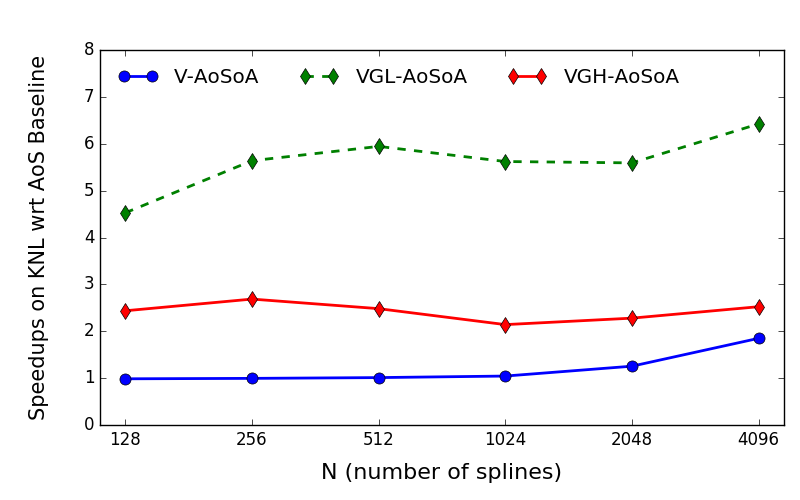}
    \caption{Normalized speedup on KNL with original AoS version as reference.}
    \label{fig:knlspeedup}
\end{figure}

Figure~\ref{fig:knlspeedup} summarizes the performance improvement on KNL for 
all the three functions \V, $\VGL$ and $\VGH$ with the AoSoA transformation.
Speedups are computed keeping the current AoS implementation 
in C/C++ in QMCPACK distribution as reference.  
AoS-to-SoA transformation does not apply to $\V$ and 
it only gets speedup with tiling. 
Speedup for \VGL\; includes performance gain from 
the basic optimizations stated earlier, 
which are in addition to the AoSoA optimization; 
and provide 
greater overall speedup. 
Our optimizations boost the throughput by 1.85x(\V), 6.4x(\VGL) 
and 2.5x(\VGH) on a node at $N=4096$.

In QMCPACK, each thread owns a \code{ParticleSet}
object and 
can benefit from keeping it in cache during the entire run. 
This AoSoA approach will
facilitate similar cache-blocking efficiency in the full application. 
The optimal tile size $N_b$ is independent of problem size $N$ for a
particular cache-based architecture and 
can be estimated based on the cache sizes present and their sharing properties.
For the production runs, $N_b$ can be tuned once for each architecture using miniQMC,
making the new algorithm cache-oblivious for any $N$ and $N_g$.  

\subsection{Exploiting Parallelism within Walker Update}
\noindent
Extending parallelism beyond the parallelization over walkers, 
is a necessary step to
scale out to more nodes, reducing the time-to-solution and the memory footprints.  
We use $n_{th}$ threads for
each walker update and reduce the number of walkers on a node by
the same factor. This allows us to scale the same problem size from one node
to $n_{th}$ nodes.  
This is well justified since the MPI efficiency remains
perfect up to 1000s of nodes on multiple HPC systems~\cite{qmcpackhybrid}.  

\begin{figure}[b]
    \centering
    \includegraphics[scale=0.35]{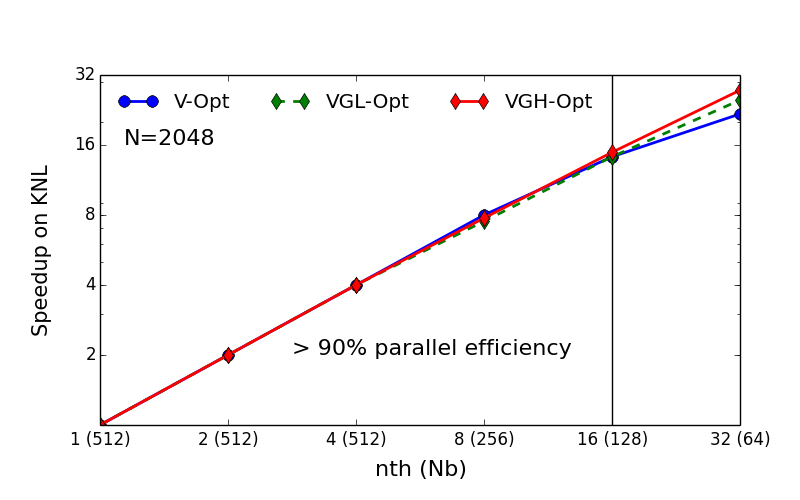}
    \caption{Scaling with respect to the number of threads per walker on KNL.
    The tile sizes $N_b$ are chosen to have sufficient number of tiles for $n_{th}$}
    \label{fig:2048s_64a_threads}
\end{figure}

Performance with the nested threading again heavily depends on 
cache system properties.
The optimal tile sizes for \xeonRR and BG/Q 
processors are determined mainly by 
the input working set size $4 n_{th} N_g N_b$ (bytes).
This increases with $n_{th}$, reducing the optimal tile
size by the same factor and limiting the scalability. 
For $N=2048$, scaling on these platforms are limited to only 2 threads
per walker for close to 80\% parallel efficiency.  
For KNC and KNL, the optimal tile size is determined by the 
output working set size; for \VGH\; $40 n_{th} N_w N_b$ bytes.
This remains constant due to the decrease in $N_w$ by the same 
$n_{th}$ factor; keeping the optimal tile size same.
This allows for strong scaling with nearly ideal parallel efficiency 
till $n_{th}=N/N_b$ for the optimal $N_b$. 
Beyond that, the scaling depends on the performance with
smaller non-optimal tile sizes. 
For $N=2048$, scaling on KNC is limited to 8 threads per walker
corresponding to $N_b=256$ for close to 80\%
parallel efficiency. 

%
%

Figure~\ref{fig:2048s_64a_threads} presents the speedup of the three B-spline
kernels in miniQMC on KNL at $N$=2048 with respect to $n_{th}$. The total number of walkers per node and memory usage are
reduced by the same $n_{th}$ factor. 
The optimal configuration of the AoSoA version is used
as the reference for computing speedups with $N_b$=512, $N_w$=256 and $n_{th}$=1.
The parallel efficiency for $n_{th}$=16 is greater than 90\%, even though
$N_b$=128 is smaller than the optimal tile size. For bigger problem size of
$N=4096$, we can expect the same parallel efficiency at 32 threads.

\section{Performance Summary and Roofline analysis}
\noindent
Our optimization processes enhance the utilization of cache and memory
bandwidth on all the four distinct cache-based architectures, discussed here. 
The relative impact of each
optimization step, varies depending upon the memory bandwidth and cache subsystem
properties on each platform, as summarized in Table~\ref{tab:speedup}.  The AoS-to-SoA data
layout conversion (Opt A) significantly raises the performance for all of them.
The same is true on GPUs~\cite{qmcpackGPU}.  The cache levels, sizes and
sharing properties play critical roles for the AoSoA transformation (Opt B).
The performance on the systems with shared LLC (BDW and BG/Q) is the best when a
subblock of the B-spline table can fit into LLC. On \xphiTM (co)processors,
AoSoA is an effective cache-blocking method and helps keep the output arrays
in cache for the reduction operations, 
allowing sustained performance for all problem sizes.  Finally,
nested threading (Opt C) is used to exploit a new parallelism over the AoSoA
objects. KNL benefits the most from Opt C and provides strong scaling of 
up to 14x with 16 threads on a node with roughly $90\%$ parallel efficiency.
Users can tune 
$n_{th}$ and the number of nodes to use, to optimize their productivity,
either to maximize the number of simulations per day (throughput) or minimize
the time-to-solution.



\begin{table}
\caption{Speedups of time of $N$=2048.
Optimization steps are A (AoS-to-SoA), B (AoSoA) and C (nested threading).
The speedups for C include the strong-scaling factor $n_{th}$.
}
\label{tab:speedup}
\centering
\begin{tabular}{|c|c|r|r|r|r|}
\hline
\multicolumn{2}{|c|}{} & BDW & KNC & KNL & BG/Q\\
\hline
\V & A/B  &  2.0       &  1.2       &  1.3       &  1.3\\
\cline{2-6}
 & C    & 3.4   &  5.9      &  18.7       &  2.0\\ 
\hline
\VGL & A  &  4.2  &  4.0  &  5.1  &  7.4 \\
\cline{2-6}
& B  &  10.2 &  5.7  &  5.6  &  9.5 \\
\cline{2-6}
& C &  17.2  &  42.1  &  80.6  & 15.8 \\
\hline
\VGH & A  & 1.7&  2.6 &  1.7 &  1.9\\
\cline{2-6}
& B  & 3.7&  5.2 &  2.3 &  2.7\\
\cline{2-6}
& C   & 6.4&  35.2 &  33.1 & 5.2\\
\hline
\multicolumn{2}{|c|}{$n_{th}$$(N_b)$ for C}  &  2(32)     &  8(256)    &  16(128)    &  2(32)\\
\hline
\end{tabular}
\end{table}

\begin{figure*}[th]
\centering
\begin{minipage}{.45\textwidth}
\includegraphics[scale=0.35]{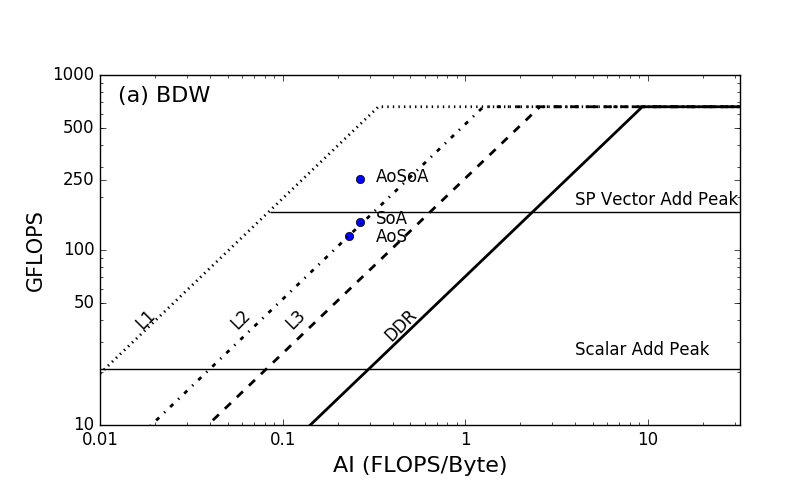}
\end{minipage}
\begin{minipage}{.45\textwidth}
\includegraphics[scale=0.35]{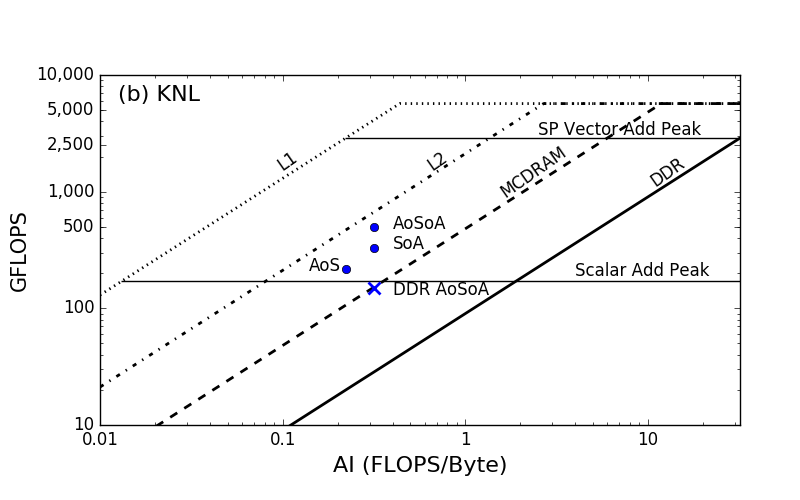}
\end{minipage}
\caption{VGH roofline performance model for $N$=2048. Circles denote
  GFLOPS at the cache-aware AI and \texttt{X} (b) the best performance (AoSoA) 
on DDR.}
    \label{fig:roofline}
\end{figure*}

Figure~\ref{fig:roofline} presents roofline performance 
analysis\cite{roofline}\cite{cacheRoofline}
of \VGH\; at various optimization steps for $N$=2048 splines.  Intel\registered
Advisor 2017~\cite{advisor} is used to compute GFLOPS and \textit{cache-aware}
arithmetic intensity (AI), FLOPS per Byte, with static and run-time analysis.
The rooflines based on the measured bounds on BDW and KNL are shown for
guidance.  

In all cases, the bytes transferred from the main memory are the same, 64$N$
reads and 10$N$ writes, and the difference in AI reflects the SIMD efficiency
and cache reuse. The AoS-to-SoA transformation increases the AI as well as
GFLOPS as we eliminate the inefficient gather and scatter 
instructions for the strided
access to the outputs.  The AoSoA transformation does not affect the AIs but
increases the performance with the optimal tile sizes through the increased
cache locality.  Higher bandwidth available with MCDRAM on KNL is critical for
higher performance, as indicated by the best 150 GFLOPS obtained on DDR
with the AoSoA version.


In addition to optimizing B-spline orbital evaluations, we are
applying the techniques described here such as SoA and AoSoA data layout 
transformations and nested threading,
for the other compute intensive kernels of miniQMC and QMCPACK.
Combining these optimizations, we get higher than 4.5x speedup 
of full miniQMC for a range of problem sizes on KNL and BDW systems.
Our goal is to transfer these performance gains 
to QMCPACK with minimal code changes.
Our SoA container classes developed in object oriented C++ 
and optimized compute kernels will be directly called from 
QMCPACK driver code. And, with the use of operator overloading feature in C++, 
non performance critical parts of QMCPACK require minimal
changes for the transition.

%% file: conclusion.tex
\section{Conclusion and Future Work}
\noindent
We presented node-level optimizations of B-spline based SPO evaluations 
widely used in QMC,
on multi/many-core shared memory processors 
and nested parallelism implementation, enabling strong scaling to reduce memory 
usage and time to solution.
The techniques described in this work, such as 
SoA data layout transformation, cache-blocking 
and threading over the AoSoA objects
are applicable to a broad range of algorithms and applications and must be
taken into account for optimizations and code modernization.  
We are applying
them in full QMCPACK to deliver the performance obtained in miniQMC
and enable breakthrough QMC studies.  
We demonstrate the impact of our optimization techniques on 
four distinct cache-coherent architectures and 
on a range of problem sizes, used in 
current day simulations to the highly futuristic ones. 
Our methods to
improve the performance 
enable performance portable solutions in QMCPACK and other
QMC applications  on current and future cache-coherent architectures.

%% file: draft_spline.bbl
\begin{thebibliography}{10}
\providecommand{\url}[1]{#1}
\csname url@samestyle\endcsname
\providecommand{\newblock}{\relax}
\providecommand{\bibinfo}[2]{#2}
\providecommand{\BIBentrySTDinterwordspacing}{\spaceskip=0pt\relax}
\providecommand{\BIBentryALTinterwordstretchfactor}{4}
\providecommand{\BIBentryALTinterwordspacing}{\spaceskip=\fontdimen2\font plus
\BIBentryALTinterwordstretchfactor\fontdimen3\font minus
  \fontdimen4\font\relax}
\providecommand{\BIBforeignlanguage}[2]{{%
\expandafter\ifx\csname l@#1\endcsname\relax
\typeout{** WARNING: IEEEtran.bst: No hyphenation pattern has been}%
\typeout{** loaded for the language `#1'. Using the pattern for}%
\typeout{** the default language instead.}%
\else
\language=\csname l@#1\endcsname
\fi
#2}}
\providecommand{\BIBdecl}{\relax}
\BIBdecl

\bibitem{ellipticene}
A.~Benali, L.~Shulenburger, N.~A. Romero, J.~Kim, and O.~A. von Lilienfeld,
  ``Application of diffusion monte carlo to materials dominated by van der
  waals interactions,'' \emph{Journal of chemical theory and computation},
  vol.~10, no.~8, pp. 3417--3422, 2014.

\bibitem{graphiteQMC}
H.~Shin, S.~Kang, J.~Koo, H.~Lee, J.~Kim, and Y.~Kwon, ``Cohesion energetics of
  carbon allotropes: Quantum monte carlo study,'' \emph{The Journal of chemical
  physics}, vol. 140, no.~11, p. 114702, 2014.

\bibitem{intercolationQMC}
P.~Ganesh, J.~Kim, C.~Park, M.~Yoon, F.~A. Reboredo, and P.~R. Kent, ``Binding
  and diffusion of lithium in graphite: quantum monte carlo benchmarks and
  validation of van der waals density functional methods,'' \emph{Journal of
  Chemical Theory and Computation}, vol.~10, no.~12, pp. 5318--5323, 2014.

\bibitem{phosphorus}
L.~Shulenburger, A.~D. Baczewski, Z.~Zhu, J.~Guan, and D.~Tomanek, ``The nature
  of the interlayer interaction in bulk and few-layer phosphorus,'' \emph{Nano
  letters}, vol.~15, no.~12, pp. 8170--8175, 2015.

\bibitem{zinc}
J.~A. Santana, J.~T. Krogel, J.~Kim, P.~R. Kent, and F.~A. Reboredo,
  ``Structural stability and defect energetics of zno from diffusion quantum
  monte carlo,'' \emph{The Journal of chemical physics}, vol. 142, no.~16, p.
  164705, 2015.

\bibitem{cuprate}
K.~Foyevtsova, J.~T. Krogel, J.~Kim, P.~Kent, E.~Dagotto, and F.~A. Reboredo,
  ``Ab initio quantum monte carlo calculations of spin superexchange in
  cuprates: The benchmarking case of ca 2 cuo 3,'' \emph{Physical Review X},
  vol.~4, no.~3, p. 031003, 2014.

\bibitem{benchmark}
L.~Shulenburger and T.~R. Mattsson, ``Quantum monte carlo applied to solids,''
  \emph{Physical Review B}, vol.~88, no.~24, p. 245117, 2013.

\bibitem{knlhc15}
\BIBentryALTinterwordspacing
``Knights landing (knl): 2nd generation intel\registered xeon phi\trademark
  processor.'' [Online]. Available: \url{http://www.hotchips.org/}
\BIBentrySTDinterwordspacing

\bibitem{bgq}
\BIBentryALTinterwordspacing
``Top 500, june 2013.'' [Online]. Available:
  \url{http://www.top500.org/lists/2013/06/}
\BIBentrySTDinterwordspacing

\bibitem{deserno1998mesh}
M.~Deserno and C.~Holm, ``How to mesh up ewald sums. i. a theoretical and
  numerical comparison of various particle mesh routines,'' \emph{The Journal
  of chemical physics}, vol. 109, no.~18, pp. 7678--7693, 1998.

\bibitem{mklCite}
\BIBentryALTinterwordspacing
``Intel\registered math kernel library (intel\registered mkl).'' [Online].
  Available: \url{https://software.intel.com/en-us/intel-mkl}
\BIBentrySTDinterwordspacing

\bibitem{qmcpackhybrid}
\BIBentryALTinterwordspacing
J.~Kim, K.~P. Esler, J.~McMinis, M.~A. Morales, B.~K. Clark, L.~Shulenburger,
  and D.~M. Ceperley, ``Hybrid algorithms in quantum monte carlo,''
  \emph{Journal of Physics: Conference Series}, vol. 402, no.~1, p. 012008,
  2012. [Online]. Available:
  \url{http://stacks.iop.org/1742-6596/402/i=1/a=012008}
\BIBentrySTDinterwordspacing

\bibitem{einspline}
\BIBentryALTinterwordspacing
K.~P. Esler, ``Einspline b-spline library.'' [Online]. Available:
  \url{http://einspline.sf.net}
\BIBentrySTDinterwordspacing

\bibitem{qmcpackGA}
\BIBentryALTinterwordspacing
Q.~Niu, J.~Dinan, S.~Tirukkovalur, A.~Benali, J.~Kim, L.~Mitas, L.~Wagner, and
  P.~Sadayappan, ``Global-view coefficients: a data management solution for
  parallel quantum monte carlo applications,'' \emph{Concurrency and
  Computation: Practice and Experience}, pp. n/a--n/a, 2016, cpe.3748.
  [Online]. Available: \url{http://dx.doi.org/10.1002/cpe.3748}
\BIBentrySTDinterwordspacing

\bibitem{YeSC15}
\BIBentryALTinterwordspacing
Y.~Luo, A.~Benali, and V.~Morozov, ``Accelerating the b-spline evaluation in
  quantum monte carlo,'' 2015. [Online]. Available:
  \url{http://sc15.supercomputing.org/sites/all/themes/SC15images/tech_poster/tech_poster_pages/post337.html}
\BIBentrySTDinterwordspacing

\bibitem{qmcpackGPU}
\BIBentryALTinterwordspacing
K.~P. Esler, J.~Kim, L.~Shulenburger, and D.~M. Ceperley, ``Fully accelerating
  quantum monte carlo simulations of real materials on gpu clusters,''
  \emph{Computing in Science and Engineering}, vol.~14, p.~40, 2012. [Online].
  Available: \url{http://doi.ieeecomputersociety.org/10.1109/MCSE.2010.122}
\BIBentrySTDinterwordspacing

\bibitem{qmcpack}
\BIBentryALTinterwordspacing
``Qmcpack.'' [Online]. Available: \url{http://www.qmcpack.org}
\BIBentrySTDinterwordspacing

\bibitem{Pearls1.book}
J.~Jeffers and J.~Reinders, Eds., \emph{High Performance Parallelism Pearls:
  Multicore and Many-core Programming Approaches}.\hskip 1em plus 0.5em minus
  0.4em\relax Boston, MA, USA: Morgan Kaufmann Publishers Inc., 2015, vol.~1.

\bibitem{Pearls2.book}
J.~Reinders and J.~Jeffers, Eds., \emph{High Performance Parallelism Pearls:
  Multicore and Many-core Programming Approaches}.\hskip 1em plus 0.5em minus
  0.4em\relax Boston, MA, USA: Morgan Kaufmann Publishers Inc., 2015, vol.~2.

\bibitem{stdl}
\BIBentryALTinterwordspacing
``Intel\registered simd data layout templates (intel\registered sdlt).''
  [Online]. Available: \url{https://software.intel.com/en-us/node/600110}
\BIBentrySTDinterwordspacing

\bibitem{sbb}
\BIBentryALTinterwordspacing
A.~Wells, ``Simd building block.'' [Online]. Available: \url{unpublished}
\BIBentrySTDinterwordspacing

\bibitem{blips4QMC}
D.~Alf{\`{e}} and M.~J. Gillan, ``{An efficient localized basis set for quantum
  Monte Carlo calculations on condensed matter},'' \emph{Physical Review B},
  vol.~70, no.~16, p. 161101, 2004.

\bibitem{coral}
\BIBentryALTinterwordspacing
``Coral collaboration, benchmark codes.'' [Online]. Available:
  \url{http://https://asc.llnl.gov/CORAL-benchmarks/}
\BIBentrySTDinterwordspacing

\bibitem{intelcomp}
``Intel compiler options: -g -ip -restrict -unroll -o3 -qopenmp -std=c++11",
  with -march=core-avx2 (bdw), -mmic (knc) and -xmic-avx512 (knl).''

\bibitem{ibmcomp}
``Ibm compiler options: -g -o3 -qinline=auto:level=10 -qarch=qp -qsmp=omp
  -qthreaded -qstrict -qhot=level=1 -qtune=qp -qsimd=auto.''

\bibitem{vtuneCite}
\BIBentryALTinterwordspacing
``Intel\registered vtune\trademark amplifier 2016.'' [Online]. Available:
  \url{https://software.intel.com/en-us/intel-vtune-amplifier-xe}
\BIBentrySTDinterwordspacing

\bibitem{hpctoolkit}
\BIBentryALTinterwordspacing
``Hpctoolkit.'' [Online]. Available: \url{http://www.hpctoolkit.org}
\BIBentrySTDinterwordspacing

\bibitem{FFTW05}
M.~Frigo and S.~G. Johnson, ``The design and implementation of {FFTW3},''
  \emph{Proceedings of the IEEE}, vol.~93, no.~2, pp. 216--231, 2005, special
  issue on ``Program Generation, Optimization, and Platform Adaptation''.

\bibitem{roofline}
S.~Williams, A.~Waterman, and D.~Patterson, ``Roofline: An insightful visual
  performance model for floating-point programs and multicore architectures,''
  \emph{Communications of the ACM}, 2009.

\bibitem{cacheRoofline}
A.~Ilic, F.~Pratas, and L.~Sousa, ``Cache-aware roofline model: Upgrading the
  loft,'' \emph{IEEE Computer Architecture Letters}, vol.~13, no.~1, pp.
  21--24, 2014.

\bibitem{advisor}
\BIBentryALTinterwordspacing
``Intel\registered advisor 2017.'' [Online]. Available:
  \url{https://software.intel.com/en-us/articles/getting-started-with-intel-advisor-roofline-feature}
\BIBentrySTDinterwordspacing

\end{thebibliography}
